\documentclass[prl,showpacs,notitlepage,twocolumn,final,floatfix]{revtex4}
\usepackage{graphicx}
\usepackage{amsmath}

\newcommand{\erf}{\operatorname{erf}}

\renewcommand{\mid}{\mathop{|}}

\begin{document}
\abovecaptionskip=0.0ex
\belowcaptionskip=2ex
\title{Firing Time Statistics for Driven Neuron Models:
  Analytic Expressions versus Numerics}
\date{\today}
\author{Michael Schindler}
\author{Peter Talkner}
\author{Peter H\"anggi}
\affiliation{Institut f\"ur Physik, Universit\"at Augsburg,
  Universit\"atsstra\ss{}e 1, D--86135 Augsburg, Germany}

\begin{abstract}
Analytical expressions are put forward to investigate the forced
spiking activity of abstract neuron models such as the driven leaky
integrate-and-fire model. The method is valid in a wide parameter
regime beyond the restraining limits of weak driving (linear response)
and/or weak noise. The novel approximation is based on a discrete
state Markovian modeling of the full long-time dynamics with
time-dependent rates. The scheme yields excellent agreement with
numerical Langevin and Fokker-Planck simulations of the full
non-stationary dynamics, not only for the first-passage time
statistics, but also for the important interspike interval (residence
time) distribution.
\end{abstract}
\maketitle

The detailed modeling of neural behavior presents a prominent
challenge on the intriguing vista towards the understanding of
neural coding principles. The leaky integrate-and-fire (LIF)
model, whose deterministic formulation has been introduced long
ago~\cite{Lap}, likely is one of the most studied abstract neuron
models~\cite{Tuckwell}. It is characterized by marked simplicity
and lack of memory: Whenever the neuron has been excited to fire a
pulse it is reset to a predefined state. The beneficial r\^ole of
an appreciable dose of noise has proved to bestow a key part to
the interspike statistics of neurons~\cite{Tuckwell}. There exist
by now numerous studies and important generalizations of realistic
synaptic models, mostly of a numerical nature, that demonstrate
the rich behavior of the renewal firing probability, e.g.\ see in
Ref.~\cite{list,BulElsDoe96, Plesser}.

In presence of a time-dependent input-stimulation the stochastic
firing process becomes non-stationary, which in turn significantly
complicates the stochastic firing statistics. Nevertheless, the signal
transmission and its detection can exhibit a remarkable improvement
via the phenomenon of \textit{Stochastic
Resonance}~\cite{GamHaeJun98}. The dynamics of the neuronal firing
probability emerges due to a large bombardment of synaptic spike
events; consequently, it is customary to employ a diffusion
approximation for the stochastic dynamics of the membrane
potential~$x(t)$. The complexity of the  driven abstract LIF model
thus assumes the archetype, non-stationary Langevin dynamics
\begin{equation}\label{eq:dim-langevin}
  \dot x(t) = -\lambda x(t) + \mu + f(t) + \sqrt{2 D}\, \xi(t)
\end{equation}
where the process starts at a time $s$ at $x(s) = x_0$ and fires when
it reaches the threshold voltage $x=a$. Here, $f(t)$~presents a
general, time-dependent stimulus which, for example, can be chosen to be
oscillatory, and $\xi(t)$~is white Gaussian noise. The dynamics of the
process~$x(t)$ is equivalently described by a Fokker-Planck (FP)
equation for the conditional probability density~$\rho(x,t\mid x_0,s)$
in a time-dependent quadratic potential,
$U(x,t) = \lambda\bigl[x-x_\text{min}(t)\bigr]^2/2$ with
$x_\text{min}(t) = \bigl(\mu + f(t)\bigr)/\lambda$, reading
\begin{equation}\label{eq:fokker-planck}
  \partial_t\, \rho = L(t) \rho
   = \partial_x\,\bigl(U^\prime(x,t) \rho\bigr) + D\partial_x^2\,\rho\:,
\end{equation}
with the absorbing boundary and initial conditions
\begin{align}\label{eq:bc}
  \rho(a,t\mid x_0,s) &= 0\quad\text{for all $t$, $s$, and $x_0$}\\
  \label{eq:initial}
  \rho(x,s\mid x_0,s) &= \delta(x-x_0).
\end{align}
After firing the process immediately restarts at the instantaneous
minimum of the potential.

The set of eqs.~(\ref{eq:dim-langevin}--\ref{eq:initial}) defines
our starting point for obtaining the firing statistics of this driven
neuron model. This is a rather intricate problem
because the presence of non-stationarity and multiple time-scales for
driving and relaxation, in combination with the absorbing
boundary condition prohibits an analytical exact solution~\cite{RMP}.
Our main objective is, nevertheless, to develop a most accurate
analytical approximation that supersedes all prior attempts known
to us. Those attempts, in fact, all involve the use of either of the
following limiting approximation schemes such as the limit of linear
response theory (i.e.~a weak stimulus~$f(t)$) \cite{LR}, the limit of
asymptotically weak noise \cite{GamHaeJun98, WN} or the use of the
method of images which appears to present an uncontrollable
approximation for the case with $\lambda \neq 0$ \cite{BulElsDoe96,
Plesser}. A most appealing numerical approach is based on an exact
integral equation for the first-passage time density of time-dependent
Gauss-Markov processes with an absorbing boundary \cite{r}. Our scheme
detailed below yields novel analytic and tractable expressions beyond
the linear response and weak noise limit; these are limited solely by
the use of a discrete, Markovian stochastic dynamics for the
population of the attracting domain and slowly varying (in comparison
to intra-well relaxation time-scale) stimuli~$f(t)$. As demonstrated
below, this novel scheme indeed provides analytical formulae that
compare very favorably with precise numerical results of the full
dynamics in eqs.~(\ref{eq:dim-langevin},
\ref{eq:fokker-planck}--\ref{eq:initial}). Different from other
approaches, we obtain the distribution not only of the first-passage
time but also of the residence time, which is the more interesting
variable, concerning neurons.

To start, we approximate the solution to
eqs.~(\ref{eq:fokker-planck}--\ref{eq:initial}) in the regime where
the statistics of times at which the threshold is reached can be
characterized by a time-dependent firing rate~$\kappa(t)$
\cite{RMP,TALU}.

This rate then follows from a time-scale separation in the full
FP dynamics~(\ref{eq:fokker-planck}) with boundary
condition~(\ref{eq:bc}). After a few times~$\lambda^{-1}$ of the fast
relaxation the probability density $\rho(x,t)$ assumes a slowly
varying pattern that decays with the rate~$\kappa(t)$. As in the
time-independent case, this slowly varying part of~$\rho(x,t)$ can be
expressed by a product of the (normalized) instantaneous stationary
solution $\rho_0(x,t) \propto \exp \bigl\{- U(x,t)/D \bigr\}$ to the
FP equation, satisfying $L(t) \rho_0(x,t) = 0$, and a form
function~$\zeta(x,t)$. Our ansatz thus reads
\begin{equation}\label{eq:asymptotic}
  \rho(x,t\mid x_0,s)
  {\overset{\scriptscriptstyle(t{}{-}{}s){>}\lambda^{-1}}{\simeq\rule{0pt}{1.5ex}}}
  \!\!\!\zeta(x,t)\: \rho_0(x,t)\:
  \exp\Bigl(-{\textstyle\int\limits_s^t}\kappa(s^\prime)\,ds^\prime\Bigr).
\end{equation}
The initial time~$s$ enters only through the exponential factor. The
dependence of the conditional probability on the initial value~$x_0$
decays exponentially on the timescale~$\lambda^{-1}$ and therefore can
be neglected for long times.

Deep inside the attracting  well, i.e.\ for $x \ll a$ there is no
sensible difference between $\rho(x,t)$~and $\rho_0(x,t)$, and
consequently $\zeta(x,t)$~approaches one. At the absorbing boundary,
however, $\rho(a,t)$ and consequently $\zeta(a,t)$ both must vanish.
The quantitative form of $\zeta(x,t)$ in the crossover region follows
from
\begin{equation} \label{eq:Lz}
  L^+(t) \zeta(x,t) = 0,
\end{equation}
where the potential in the backward operator $L^+(t) =
-U'(x,t)\partial_x + D \partial^2_x$ can be linearized about the
threshold $x=a$. Eq.~(\ref{eq:Lz}) then yields for the form function
the result
\begin{equation} \label{eq:z}
  \zeta(x,t)= 1- \exp \Bigl\{ (x-a)\frac{U'(a,t)}{D} \Bigr\}.
\end{equation}
The rate~$\kappa(t)$ is determined by multiplying the
FP equation~(\ref{eq:fokker-planck}) in the long-time limit~(\ref{eq:asymptotic})
by the form function~$\zeta(x,t)$ and integrating over~$x$ from
$-\infty$~to the threshold voltage~$a$. In doing so, we account for
prominent \textit{finite barrier corrections}, yielding
\begin{equation}\label{eq:k1}
  \kappa(t) = - \frac{\int_{-\infty}^a
  dx\:\zeta(x,t) L(t) \zeta(x,t)
  \rho_0(x,t)} {\int_{-\infty}^a dx\: \zeta^2(x,t) \rho_0(x,t)}\;.
\end{equation}
Upon insertion of eq.~(\ref{eq:z}) for the form function one can
exactly perform the integrations and obtains for the rate
\begin{equation}\label{eq:rate}
  \kappa(t) = \lambda\; \frac{\Delta U(t)}{D}\; \frac{1 -
  \erf\bigl(\sqrt{\Delta U(t)/D}\,\bigr)}{1 - \exp(-\Delta U(t)/D)} \;,
\end{equation}
where $\Delta U(t)$ denotes the instantaneous potential height at the
threshold as seen from the minimum, and $\erf(z)$~is the error
function. For very small~$D$ an expansion of the error function leads
to the well-known \textit{weak noise} result for the time-dependent
rate \cite{GamHaeJun98,RMP}, i.e.
\begin{equation}\label{eq:a_rate}
  \kappa^\text{wn}(t) = \lambda\; \sqrt{\Delta U(t)/(\pi D)}\;
  \exp(-\Delta U(t)/D)\;.
\end{equation}

\textit{Firing time distributions}.---With the expression for the exit
rate~$\kappa(t)$ in~(\ref{eq:rate}) we can calculate the properties of
interest, namely the densities for the first-passage time and the
residence time~\cite{TD} of the attracting "integrating" state that
covers the domain $-\infty<x(t)<a$.

The first-passage time distribution is given by the negative rate of
change of probability finding the process at time~$t$ in the
"integrating" state, i.e.
\begin{align}
  g(t\mid s) &= - \partial_t \int_{-\infty}^a \rho(x,t\mid x_0,s)\,dx\\
   \label{eq:first-passage}
   &= \kappa(t)\:\exp\Bigl(-{\textstyle\int\limits_s^t}
       \kappa(s^\prime)\,ds^\prime\Bigr)\:,
\end{align}
Here, the integral over the spatial part of~$\rho(x,t)$ yields unity
and the exponential factor in~(\ref{eq:asymptotic}) is obtained. It
gives the probability for the process to stay in the "integrating"
state from time~$s$ until~$t$ without interruption.

The distribution of the residence times~$h(\tau)$, also termed the
\textit{interspike interval density}, follows as the average of the
first-passage time density over the density of resetting times~$s$,
which coincides with the firing rate~$\kappa(s)$. It thus reads:
\begin{equation}
  h(\tau) = \lim_{T \to \infty} \frac{\int_{-T}^T g(\tau+s\mid s)
  \kappa(s) ds} {\int_{-T}^T \kappa(s) ds} \label{eq:residence}\;.
\end{equation}
Equations (\ref{eq:first-passage})~and (\ref{eq:residence}), together
with the expression for the rate~(\ref{eq:rate}) constitute the main
results of this work. Their quantitative validity for an extended
parameter regime  will be checked next.
\begin{figure}[!b]%
\centering
\includegraphics{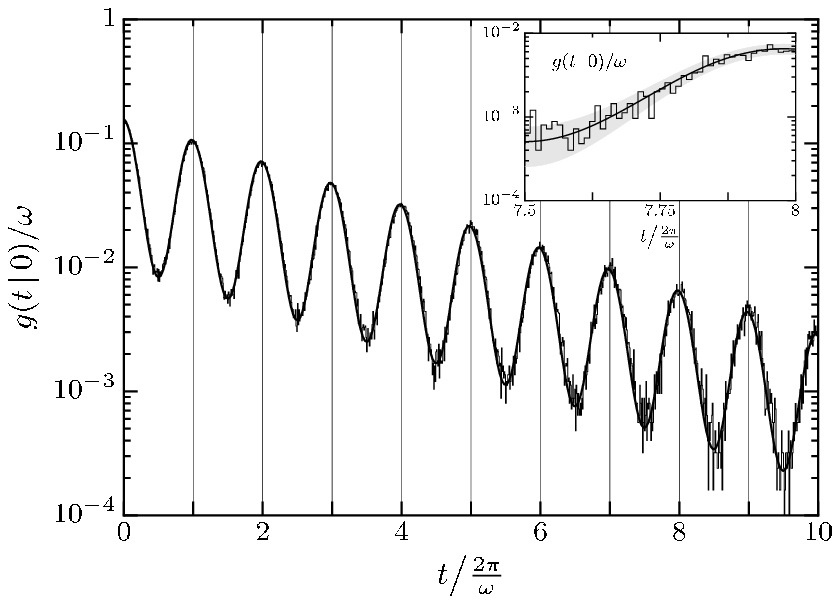}%
\caption{First-passage time density for parameters $U_+/D = 8$, $U_-/D
= 5$, $\omega=0.05$. The jagged line depicts the histogram obtained
from iterations of the Langevin equation in~(\ref{eq:langevin}). Note
that fluctuations in the histogram depend on the total number of
events and the width of the histogram bins. These fluctuations stay
completely within their expected range which is indicated as the gray
shaded area in the inset. The height of this area is twice the
expected standard deviation of the histogram levels. The solid
line shows the analytic first-passage time density from
eq.~(\ref{eq:first-passage}), with the rate used in~(\ref{eq:rate}).
Both lines are in excellent agreement.}%
\label{fig:opt}%
\includegraphics{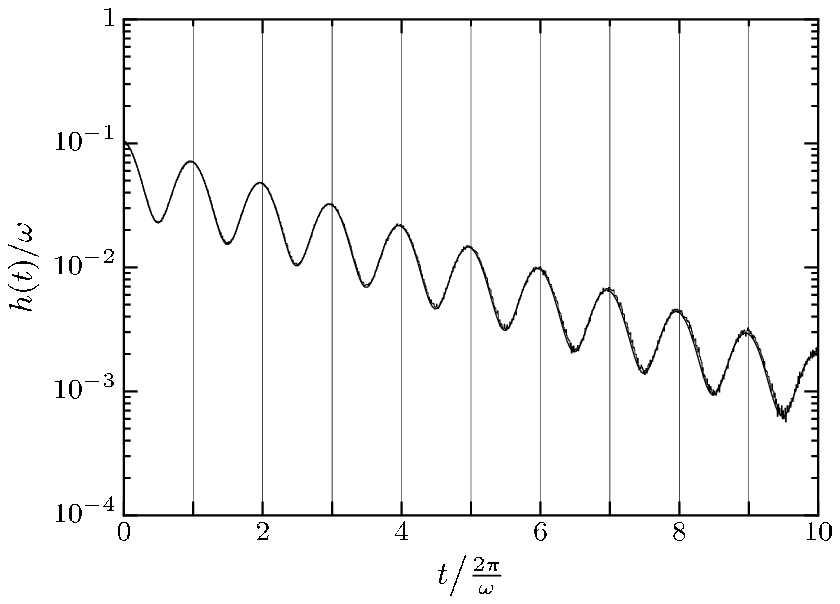}%
\caption{Residence time density \textit{vs}.~time for the same
parameters as in Fig.~\ref{fig:opt}. The jagged line shows the
histogram obtained from iterations of~(\ref{eq:langevin}). Again, the
numerics practically coincides within the line-width with the analytic
expression in~(\ref{eq:residence}) evaluated with the rate
in~(\ref{eq:rate}) (solid line).}%
\label{fig:res}%
\end{figure}
\begin{figure}[!b]%
\centering
\includegraphics{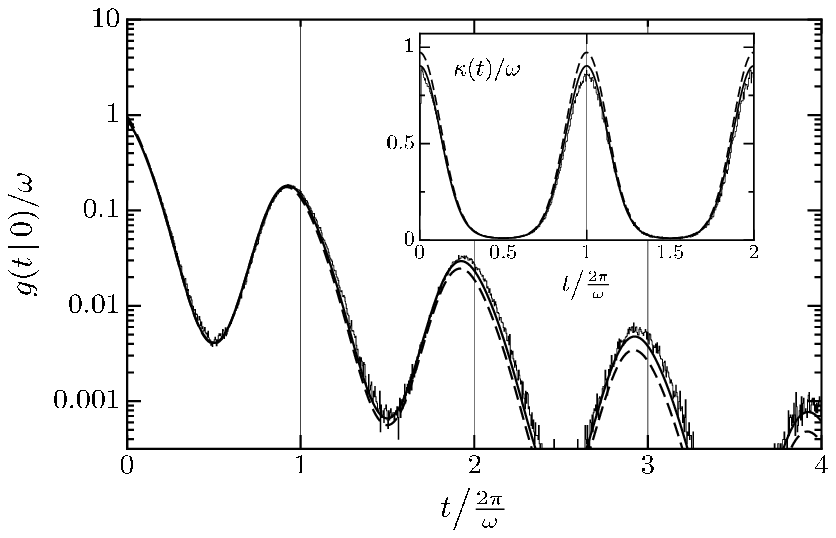}%
\caption{Testing extreme limits. First-passage time density for an
extremely small lower barrier $U_-/D = 3$. The remaining
parameters are as in Fig.~\ref{fig:opt}. Langevin simulation
results (jagged), analytical result in (\ref{eq:first-passage})
with eq.~(\ref{eq:rate}) (solid), likewise, with
eq.~(\ref{eq:a_rate}) (dashed). The inset compares the rate
$\kappa(t)$ obtained from simulations of~(\ref{eq:langevin}) with
the analytic results from~eq.~(\ref{eq:rate}) and
eq.~(\ref{eq:a_rate}), respectively.}%
\label{fig:low}%
\includegraphics{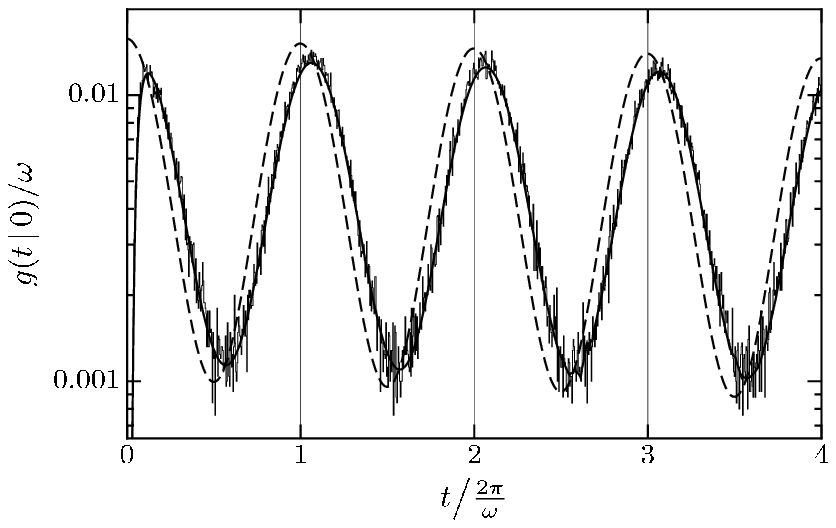}%
\caption{Testing extreme limits. Probability density of the
first-passage time for a very fast driving, $\omega=0.5$. The
other parameters are as in Fig.~\ref{fig:opt}. The analytic
approximation~(eqs.~(\ref{eq:first-passage}) and (\ref{eq:rate}),
dashed line) still depicts maxima that are approximately located
at~$1,2,\ldots$ while the numerical results are shifted to later
times. The jagged line presents the Langevin-iterations
from~(\ref{eq:langevin}). The solid curve presents within its
linewidth both the numerical solution to the FP
equation~(\ref{eq:fokker-planck}) and to the integral equation
from~\cite{r}.}%
\label{fig:fast}%
\end{figure}

\textit{Numerical comparison}.---We have employed three methods for
the numerical analysis. A first one is based on the Langevin
equation~(\ref{eq:dim-langevin}) where the position~$x(t)$ is updated
sequentially. For the second we have solved the FP
equation~(\ref{eq:fokker-planck}) numerically, using a Chebychev
collocation method to reduce the problem to a coupled system of
ordinary differential equations, see also~\cite{Lehmann}. The third
method solves an integral equation for the first-passage time
probability density and is described in~\cite{r}. All three methods
have provided practically identical results.

In the prior stimulating work~\cite{BulElsDoe96} it has been left open
in what relevant parameter regime the employed approximation possesses
validity~\cite{Plesser}. Here, using a periodic modulation
$f(t)=A\cos(\omega t)$ we like to determine a minimal set of relevant
parameters that can be taken for comparison with physiological
measurements. The most general Langevin
equation~(\ref{eq:dim-langevin}) considered has seven constant
parameters, including the threshold~$a$ and the reset-position~$x_0$.
Three of them, $(\lambda,\mu,a)$, can be chosen to be $(1,0,1)$ by
transforming to dimensionless coordinates, using the
time-unit~$\smash{\lambda^{-1}}$, the space-unit $(a-\mu/\lambda)$, and
the coordinate-origin at~$\mu/\lambda$. The resulting parameters thus
read (the bars indicate dimensionless coordinates)
\begin{equation}
\bar A = A /(\lambda a - \mu),\;
  \bar\omega = \omega/\lambda, \;
  \bar D = D/\bigl(\lambda(a - \mu\lambda)^2\bigr)\:.
\end{equation}
For our purposes it is advantageous to use instead the equivalent set
\begin{equation}
 \bar U_+/\bar D,\;
  \bar U_-/\bar D,\text{ and }
  \bar\omega,
\end{equation}
where~$\bar U_+$ and $\bar U_-$ are the maximum and the minimum
of~$\bar U(1,\bar t)$ during a whole period of modulation. These so
chosen parameters have the benefit that they provide an on-hand
estimate for the validity of our approximations and can be evaluated
directly from~(\ref{eq:dim-langevin}).

The equation we have used in our simulations thus reads (omitting
the overbars)
\begin{equation}\label{eq:langevin}
  \dot{x}(t) = - x(t) + A\,\cos(\omega t) +
  \sqrt{2 D\,}\: \xi(t)
\end{equation}
with the threshold located at $\thickmuskip=\thinmuskip x=1$. For
obtaining the residence time, $x$~has been reset into the minimum
of~$U(x,t)$ immediately after firing. The Figs.~\ref{fig:opt}
and~\ref{fig:res} depict the probability densities of the
first-passage and the residence times, respectively. The residence
time distribution exhibits a less pronounced modulation in comparison
with the first-passage time distribution. Both analytical
expressions in (\ref{eq:first-passage})~and (\ref{eq:residence})
compare very favorably with the numerical results obtained by
iteration of the Langevin equation~(\ref{eq:langevin}). The remaining
deviations in the two figures are of purely statistical nature (see
the inset in Fig.~\ref{fig:opt}) and can be diminished further by
increasing the number of events in the simulations.

In order to further test the range of validity of our
novel approximation scheme we -- on purpose -- have chosen extreme
values for the lower barrier height $U_-/D$ and angular driving
frequency $\omega$, respectively, see Figs.~\ref{fig:low}
and~\ref{fig:fast}. Here, deviations from the numerical results
are not the result of statistics but are systematic. For the low
potential barrier in Fig.~\ref{fig:low}, the time-scales in the
process are not separated sufficiently. The fast intra-well
fluctuations begin to influence the behavior of the modulated
firing dynamics. Moreover, the difference between the moderate
noise result for the time-dependent rate~$\kappa(t)$ in
(\ref{eq:rate}) and its weak noise approximation in
(\ref{eq:a_rate}) becomes visibly increased, as expected.
Figure~\ref{fig:fast} depicts the other extreme situation with a
modulation time-scale that is not slow enough. Because the system
cannot follow the driving instantaneously we find a shift in the
maxima of the first-passage time density. This shift is not
reproduced by our approximation in (\ref{eq:asymptotic})~and
(\ref{eq:first-passage});
nevertheless, our scheme yields amazingly good results even within
this extreme parameter regime. The results based on the numerical
evaluation of the FP equation~(\ref{eq:fokker-planck}) and the
integral equation~\cite{r} virtually collapse into one curve and
perfectly coincide with the Langevin simulations. The same result was
obtained for the other parameter values.

\textit{Conclusions}.---The precise theoretical modeling of the
neuronal spiking activity under external time-dependent driving
presents a challenge of considerable importance in neurophysiology
and physics. Due to the presence of non-stationarity, absorbing
boundary conditions of the underlying first-passage problem and
differing time-scales the task of obtaining reliable analytical
estimates for the firing statistics is anything but trivial. By
reference to a \textit{discrete} Markovian
dynamics for the corresponding full space-continuous
stochastic dynamics we succeeded in obtaining analytical
approximations for the time-dependent first-passage time and the
residence time statistics that are valid beyond the restraining limits
of linear response and asymptotically weak noise. We have tested our
findings for the case of a periodically driven LIF model.
The obtained agreement with precise numerical simulations of
either the Langevin type
in~(\ref{eq:langevin}) or, equivalently, of the FP type
in~(\ref{eq:fokker-planck}) turns out to be very good. Our method
is not restricted to an oscillatory forcing but applies as well to
arbitrary drive functions~$f(t)$ such as an exponentially decaying
drive (e.g. simulating a decaying threshold). Our scheme even
yields good results in extreme parameter regimes where
agreement cannot be expected \textit{a priori}.

Our method, primarily aimed at describing first-passage time and
residence time probabilities of driven dynamical systems, is also
readily extended to more realistic neuron models such as e.g.\
the two-dimensional driven FitzHugh-Nagumo model~\cite{FN} for
neuronal spiking activity, whose multiple attractors may be considered
as discrete states.
Likewise, the scheme can also be employed to study yet other
time-dependent switching dynamics and synchronization phenomena
such as the paradigm of \textit{Stochastic
Resonance}~\cite{GamHaeJun98} and discrete or continuous
Brownian motor transport~\cite{BM}.

This work has been supported by the Deutsche Forschungsgemeinschaft
via Projects No.~HA1517/13-4 and SFB-486, Projects No.~A10 and No.~B13.


\begin{thebibliography}{99}
\bibitem{Lap}
  L.~Lapicque, J.~Physiol. (Paris) \textbf{9}, 620 (1907).
\bibitem{Tuckwell}
  H.~C. Tuckwell, \textit{Stochastic Processes in the Neurosciences}, (SIAM, Philadelphia, 1989).
\bibitem{list}
  B.~W. Knight, J.~Gen. Phys. \textbf{59}, 734 (1972);
  M.~Stemmler, Network-Comp. Neural \textbf{7}, 687 (1996);
  P.~Lansky, Phys. Rev.~E \textbf{55}, 2040 (1997);
  M.~T. Giraudo and L.~Sacerdote, BioSystems \textbf{48}, 77 (1998);
  T.~Shimokawa \textit{et al.}, Phys. Rev.~E \textbf{59}, 3461 (1999);
  H.~E. Plesser and T.~Geisel, Phys. Rev.~E \textbf{59}, 7008 (1999);
  M.~J. Chacron \textit{et al.}, Phys. Rev. Lett. \textbf{85} 1576 (2000);
  N.~Brunel \textit{et al.}, Phys. Rev. Lett. \textbf{86}, 2186 (2001);
  L.~Sacerdote and P.~Lansky, BioSystems \textbf{67}, 213 (2002);
  J.~W. Middleton \textit{et al.}, Phys. Rev.~E \textbf{68}, 021920 (2003);
  B.~Lindner, A.~Longtin and A.~Bulsara, Neural Comp. \textbf{15}, 1761 (2003).
\bibitem{BulElsDoe96}
  A.~R. Bulsara, T.~C. Elston, C.~R. Doering, S.~B. Lowen, and K.~Lindenberg, Phys. Rev.~E \textbf{53}, 3958 (1996).
\bibitem{Plesser}
  H.~E. Plesser and W.~Gerstner, Neurocomputing \textbf{32--33}, 219 (2000);
  H.~E. Plesser and T.~Geisel, Phys. Rev.~E \textbf{59}, 7008 (1999).
\bibitem{GamHaeJun98}
  L.~Gammaitoni, P.~H\"anggi, P.~Jung, and F.~Marchesoni, Rev. Mod. Phys. \textbf{70}, 223 (1998).
\bibitem{RMP}
  P. H\"anggi, P. Talkner, and M. Borkovec, Rev. Mod. Phys. \textbf{62}, 251 (1990).
\bibitem{LR}
  B.~Lindner and L.~Schimansky-Geier, Phys. Rev. Lett. \textbf{86}, 2934 (2001);
  N.~Fourcaud and N.~Brunel, Neural Comp. \textbf{14}, 2057 (2002);
  B.~Lindner \textit{et al.}, Phys. Rep. \textbf{392}, 321 (2004).
\bibitem{WN}
  P.~Jung and P.~H\"anggi, Phys. Rev.~A \textbf{44}, 8032 (1991);
  V.~A. Shneidman \textit{et al.}, Phys. Rev. Lett. \textbf{72}, 2682 (1994);
  N.~G. Stocks, Nuovo Cimento~D \textbf{17}, 925 (1995);
  J.~Lehmann, P.~Reimann, and P.~H\"anggi, Phys. Rev. Lett. \textbf{84}, 1639 (2000);
  A.~Nikitin, N.~G.~Stocks and A. R.~Bulsara, Phys. Rev.~E \textbf{68}, 016103 (2003);
  J.~Casado-Pascual \textit{et al.}, Phys. Rev. Lett. \textbf{91}, 210601 (2003).
\bibitem{r}
  E.~Di Nardo, A.G. Nobile, E.~Pirozzi, L.~M. Ricciardi, Adv. Appl. Probab. {\bf 33}, 453 (2001);
  A.~Buonocore, \textit{et al.}, Adv. Appl. Probab. \textbf{19}, 784 (1987).
\bibitem{TALU}
  P.~Talkner and J.~Luczka, arXiv:condmat/0307498.
\bibitem{TD}
  R.~L\"ofstedt and S.~N. Coppersmith, Phys. Rev.~E \textbf{49}, 4821 (1994);
  M.~H. Choi, R.~F. Fox, and P.~Jung, Phys. Rev.~E \textbf{57}, 6335 (1998);
  P.~Talkner, Physica~A \textbf{325}, 124 (2003).
\bibitem{Lehmann}
  J.~Lehmann, P.~Reimann, and P.~H\"anggi, Phys. Rev. E \textbf{62}, 6282 (2000).
\bibitem{FN}
  R.~A. FitzHugh, Biophys.~J. \textbf{1}, 445 (1961);
  J.~Nagumo, \textit{et al.}, Proc. IRE \textbf{50}, 2061 (1962).
\bibitem{BM}
  R.~D. Astumian and P.~H\"anggi, Phys. Today \textbf{55} No.11, 33 (2002);
  P.~Reimann, Phys. Rep. \textbf{361}, 57 (2002).
\end{thebibliography}
\end{document}